\documentclass[acmsmall,screen]{acmart}

\usepackage{amsmath,amsfonts}
\usepackage{textcomp}
\usepackage{xcolor}
\usepackage{listings} %
\usepackage{wrapfig}
\usepackage[english]{babel}
\usepackage[figuresright]{rotating}
\usepackage[flushleft]{threeparttable}

\usepackage{multirow}
\usepackage{booktabs}
\usepackage{bbm}
\usepackage{array}
\usepackage{graphicx} %
\usepackage{caption}
\usepackage{float}
\usepackage{subcaption}
\usepackage{fancybox}

\usepackage{supertabular,booktabs}
\usepackage{soul}
\soulregister\cite7 %
\soulregister\citep7 %
\soulregister\citet7 %
\soulregister\ref7 %
\soulregister\pageref7 %

\usepackage[normalem]{ulem}

\newcommand{\delete}[1]{}

\newcommand{\newdelete}[1]{}

\usepackage{url}
\usepackage{hyperref}
\hypersetup{
    colorlinks=true,
    linkcolor=blue,
    filecolor=magenta,      
    urlcolor=cyan,
}

\usepackage{CJKutf8}
\usepackage{enumitem}

\usepackage[most]{tcolorbox}

\newcommand{\finding}[2]{
\begin{center}
\begin{tcolorbox}[leftrule=0mm,toprule=0mm,bottomrule=0mm,rightrule=0mm,left=1pt,right=2pt,top=0pt,bottom=0pt,breakable]
\textbf{Finding {#1}:}
{#2}
\end{tcolorbox}
\end{center}
}

\newcommand{\summary}[1]{
\begin{center}
\begin{tcolorbox}[leftrule=0mm,toprule=0mm,bottomrule=0mm,rightrule=0mm,left=1pt,right=2pt,top=0pt,bottom=0pt,breakable]
\textbf{Summary:}
{#1}
\end{tcolorbox}
\end{center}
}

\AtBeginDocument{%
  \providecommand\BibTeX{{%
    \normalfont B\kern-0.5em{\scshape i\kern-0.25em b}\kern-0.8em\TeX}}}

\setcopyright{acmcopyright}
\copyrightyear{2025}
\acmYear{2025}
\acmDOI{XXXXXXX.XXXXXXX}

\acmJournal{PACMSE}
\acmVolume{0}
\acmNumber{ISSTA}
\acmArticle{0}
\acmMonth{0}

\acmConference[ISSTA 2025]{ACM SIGSOFT International Symposium on Software Testing and Analysis}{25-28 June, 2025}{Trondheim, Norway}

\begin{document}

\title{A Large-scale Empirical Study on Fine-tuning Large Language Models for Unit Testing}

\author{Ye Shang} 
\orcid{0009-0000-8699-8075}
\authornote{Both authors contributed equally to this research.}
\email{201250032@smail.nju.edu.cn}

\author{Quanjun Zhang} 
\orcid{0000-0002-2495-3805}
\email{quanjun.zhang@smail.nju.edu.cn}
\authornotemark[1]
\affiliation{
  \institution{State Key Laboratory for Novel Software Technology, Nanjing University}
  \city{Nanjing}
  \state{Jiangsu}
  \country{China}
  \postcode{210093}
}

\author{Chunrong Fang} 
\orcid{0000-0002-9930-7111}
\email{fangchunrong@nju.edu.cn}
\authornote{Chunrong Fang is the corresponding author.}
\affiliation{
  \institution{State Key Laboratory for Novel Software Technology, Nanjing University}
  \city{Nanjing}
  \state{Jiangsu}
  \country{China}
  \postcode{210093}
}

\author{Siqi Gu} 
\orcid{0000-0001-5514-6734}
\email{siqi.gu@smail.nju.edu.cn}
\affiliation{
  \institution{State Key Laboratory for Novel Software Technology, Nanjing University}
  \city{Nanjing}
  \state{Jiangsu}
  \country{China}
  \postcode{210093}
}

\author{Jianyi Zhou}
\email{zhoujianyi2@huawei.com}
\orcid{0000-0002-4867-5416}
\affiliation{
  \institution{Huawei Cloud Computing Technologies Co., Ltd.}
  \city{Beijing}
  \country{China}
  \postcode{100085}
}

\author{Zhenyu Chen}
\email{zychen@nju.edu.cn}
\orcid{0000-0002-9592-7022}

\affiliation{
  \institution{State Key Laboratory for Novel Software Technology, Nanjing University}
  \city{Nanjing}
  \state{Jiangsu}
  \country{China}
  \postcode{210093}
}

\begin{abstract}

Unit testing plays a pivotal role in software development, improving software quality and reliability. However, generating effective test cases manually is time-consuming, prompting interest in unit testing research. 
Recently, Large Language Models (LLMs) have shown potential in various unit testing tasks, including test generation, assertion generation, and test evolution, but existing studies are limited in scope and lack a systematic evaluation of the effectiveness of LLMs. 

To bridge this gap, we present a large-scale empirical study on fine-tuning LLMs for unit testing.
Our study involves three unit testing tasks, five benchmarks, eight evaluation metrics, and 37 popular LLMs across various architectures and sizes, consuming over 3,000 NVIDIA A100 GPU hours.
We focus on three key research questions: (1) the performance of LLMs compared to state-of-the-art methods, (2) the impact of different factors on LLM performance, and (3) the effectiveness of fine-tuning versus prompt engineering. 
Our findings reveal that LLMs outperform existing state-of-the-art approaches on all three unit testing tasks across nearly all metrics, highlighting the potential of fine-tuning LLMs in unit testing tasks. 
Furthermore, large-scale, decoder-only models achieve the best results across tasks, while encoder-decoder models perform better under the same parameter scale. 
Additionally, the comparison of the performance between fine-tuning and prompt engineering approaches reveals the considerable potential capability of the prompt engineering approach in unit testing tasks. 
We then discuss the concerned issues on the test generation task, including data leakage issues, bug detection capabilities, and metrics comparisons. 
Finally, we further pinpoint carious practical guidelines for LLM-based approaches to unit testing tasks in the near future.
Overall, our work demonstrates the promising future of fine-tuning LLMs on unit testing tasks and reduces the manual efforts of unit testing experts in practical scenarios.

\end{abstract}



\maketitle

\sloppy
\section{Introduction}
\label{sec:introduction}

Unit testing is a crucial part of the software development lifecycle, significantly enhancing software quality and reliability~\cite{wang_software_2024,zhang2022test}.
It involves the systematic execution of software to detect potential bugs, verify expected behaviors, and ensure compliance with requirements, thereby enhancing developers' efficiency in development and debugging.
However, constructing effective test cases manually is both challenging and time-intensive, with studies indicating that developers often spend over 15\% of their time on test generation~\cite{daka_survey_2014}.
As a result, automated test generation has gained substantial interest among developers and researchers. Numerous approaches have been proposed to reduce manual effort, including symbolic execution testing~\cite{chipounov_s2e_2011, cadar_symbolic_2011}, model-based testing~\cite{dalal_modelbased_1999, offutt_generating_1999}, random testing~\cite{pacheco_randoop_2007, ma_grt_2015}, and search-based testing~\cite{fraser_evosuite_2011, baresi_testful_2010}.

A unit test typically includes two components: (1) a test prefix, i.e., a sequence of statements that manipulate the unit under test to a specific state, and (2) a test assertion, i.e., an oracle that defines the conditions that should be satisfied in that state~\cite{dinella_toga_2022}. 
In the literature, there are mainly three unit testing tasks based on different perspectives: test generation, assertion generation, and test evolution. 
First, the test generation task takes a focal method (i.e., the method under test ) as input and generates a unit test that includes both a test prefix and an assertion to ensure the focal method functions correctly~\cite{tufano_unit_2021}.
Second, the assertion generation task focuses on generating meaningful assert statements for a given focal method and a test prefix to capture potential faults effectively.
Third, the test evolution task attempts to generate an updated unit test consisting of both prefixes and assertions while incorporating additional inputs, including the original focal method, its unit test, and the updated focal method~\cite{hu_identify_2023}.
The test evolution task is usually conducted during software evolution, where the focal method adapts to meet new requirements or resolve issues, requiring unit tests to co-evolve to maintain software quality. 
Overall, these three tasks, i.e., test generation, assertion generation, and test evolution, represent distinct aspects of unit testing across various levels of granularity and scenarios, offering a comprehensive view of unit testing.

\textbf{This Paper.}
Recently, Large Language Models (LLMs) have demonstrated impressive performance on various code-related tasks such as code generation~\cite{li_starcoder_2023, wei_magicoder_2024}, code summarization~\cite{sun_prompt_2023, ahmed_automatic_2024}, and program repair~\cite{ zhang2024systematic,zhang2023survey}. 
In the domain of unit testing, several studies have explored LLM-based approaches with promising results. 
For instance, A3Test~\cite{alagarsamy_a3test_2023} fine-tunes a PLBART model specifically for test generation and Zhang et al.~\cite{zhang_exploring_2024} explore the capability of several LLMs for assertion generation. 
Furthermore, \textsc{Ceprot}~\cite{hu_identify_2023} leverages CodeT5, a pre-trained model designed to learn from source code, to capture semantic correlations effectively in the test evolution task. 
Despite ongoing research, the literature still lacks a systematic evaluation of LLMs in unit testing, making it difficult for researchers to assess the true capabilities of LLMs in performing various unit testing tasks.
To fill this gap, we conduct the first large-scale empirical study on LLMs for unit testing, including three unit testing tasks (i.e., test generation, assertion generation, and test evolution) across 37 LLMs with different architectures (i.e., encoder-only, encoder-decoder, and decoder-only) and size (ranging from 60 million to 16 billion parameters). 
Our study provides a comprehensive evaluation of LLMs on these common unit testing tasks, facilitating future research in this area. 
Specifically, we address the following three research questions.

\begin{itemize}[leftmargin=1em]
    \item \textbf{Performance of LLMs on Unit Testing Tasks.}  
    In RQ1, we evaluate the performance of LLMs on test generation, assertion generation, and test evolution tasks, comparing them to previous state-of-the-art approaches using two types of metrics.
    \textbf{Results:} LLMs significantly outperform several state-of-the-art approaches such as $ATLAS$ for assertion generation~\cite{watson_learning_2020}, \textsc{AthenaTest} for test generation~\cite{tufano_unit_2021}, and \textsc{Ceprot} for test evolution~\cite{hu_identify_2023}, across nearly all metrics, highlighting their potential in unit testing.

    \item \textbf{Impact of Various Factors on LLMs' Performance.}  
    In RQ2, we evaluate the effects of model series, architecture, and size on LLMs' performance. 
    \textbf{Results:} (1) large-scale LLMs consistently outperform smaller ones; (2) decoder-only models achieve the highest performance overall, with encoder-decoder models showing strength at comparable parameter sizes; (3) the CodeLlama~\cite{roziere_code_2024}, DeepSeek-Coder~\cite{guo_deepseek-coder_2024}, and CodeT5p~\cite{wang_codet5_2023+} series models are particularly promising.

    \item \textbf{Comparison between the Fine-tuning and Prompt Engineering Approaches.}
    In RQ3, we compare the effectiveness of the fine-tuning and prompt engineering approaches across three unit testing tasks. 
    \textbf{Results:} prompt engineering approach with zero-shot learning produces promising results in test generation and test evolution tasks, indicating its considerable potential capability in common unit testing tasks.

\end{itemize}

\textbf{Novelty \& Contributions.}
To sum up, the main contributions of this paper are as follows:

\begin{itemize}[leftmargin=1em]
    \item \textbf{New Dimension.} 
    We bridge the gap between LLMs and three unit testing tasks: test generation, assertion generation, and test evolution. 
    Our study not only provides a comprehensive evaluation of LLMs' performance but also highlights the potential of fine-tuning LLMs in unit testing.

    \item \textbf{Extensive Study.}
    We conduct the first large-scale empirical study on fine-tuning 37 popular LLMs in three unit testing scenarios across five benchmarks and eight metrics, utilizing over 3,000 NVIDIA A100 GPU hours.
    Our study includes (1) a systematic comparison of LLMs with state-of-the-art methods for unit testing;
    (2) an in-depth analysis of factors affecting LLMs' performance;
    and (3) a complete comparison between fine-tuning and prompt engineering approaches.

    \item \textbf{Practical Guidelines.}
    Based on our extensive findings, we provide practical guidelines to inform and guide future research and practice in applying LLM-based approaches to unit testing tasks.
\end{itemize}

\sloppy
\section{Background and Related Work}
\label{sec:background}

\subsection{Test Generation}

Unit test generation can be broadly categorized into traditional approaches and Deep Learning (DL)-based approaches. 
Traditional approaches, exemplified by tools like Evosuite~\cite{fraser_evosuite_2011} and Randoop~\cite{pacheco_randoop_2007}, utilize a series of software analysis techniques, including search-based~\cite{fraser_evosuite_2011}, random-based~\cite{pacheco_randoop_2007}, model checking~\cite{enoiu_automated_2016, gargantini_using_1999}, and symbolic execution~\cite{pasareanu_combining_2008,xie_symstra_2005}, to generate unit tests that achieve high code coverage and mutation scores.

DL-based approaches, by using pre-trained models, treat the test generation task as a neural machine translation problem, where the input primarily consists of the focal methods and the output is the generated unit tests. 
Tufano et al.~\cite{tufano_unit_2021} introduce an approach named \textsc{AthenaTest} to generate unit tests, which use a BART architecture model for a two-step training procedure on a large processed dataset. Alagarsamy et al.~\cite{alagarsamy_a3test_2023} propose an approach named A3Test. In addition to using a PLBART model, A3Test implements simple post-processing steps such as verifying naming consistency, correcting incomplete parentheses, and refining test signatures to increase the correctness rate.
More recently, with the significant impact of closed-source LLMs like ChatGPT, many researchers have explored using a prompt engineering approach to generate unit tests. Yuan et al.~\cite{yuan_no_2024} propose ChatTester, which combines an initial test generator with an iterative test refiner to produce unit tests. Chen et al.~\cite{chen_chatunitest_2024} introduce ChatUniTest, which generates unit tests under the Generation-Validation-Repair framework. 
Gu et al.\cite{gu2024testart} introduce TestART, a framework that generates high-quality unit test cases by integrating traditional template-based program repair\cite{zhang2023gamma} with the generative capabilities of LLMs within a \textit{generation-and-repair} mechanism.
Zhang et al.~\cite{zhang2024testbench} introduce TestBench, the first benchmark tailored to class-level LLM-based test case generation, involving 108 Java programs from 9 large-scale GitHub projects, three distinct prompt types and five key evaluation aspects of test cases.
Besides, some empirical studies~\cite{yang_empirical_2024, schaefer_empirical_2024} have emerged, comparing the effectiveness of different LLMs with various prompt strategies.

\subsection{Assertion Generation}

Assertion generation is primarily approached using two approaches~\cite{he2024empirical}: DL-based approaches and Information Retrieval (IR)-based approaches.
DL-based approaches treat assertion generation as a Neural Machine Translation (NMT) task. 
Watson et al.~\cite{watson_learning_2020} pioneer the use of DL-based approaches in assertion generation with their proposal of $ATLAS$. 
Later, IR-based approaches are applied to the assertion generation task as well, which use retrievers to find similar assertions based on focal-test information and then correct the retrieved assertions by replacing incorrect tokens with the correct ones based on context.
Yu et al.~\cite{yu_automated_2022} propose an IR-based method, introducing IR-based assertion retrieval ($IR_{ar}$) and retrieved-assertion adaptation ($RA_{adapt}$) approaches. $IR_{ar}$ retrieves the most similar assertion to the given focal-test, while $RA_{adapt}$ adjusts tokens based on context to refine the assertion. Yu et al. further propose an integrated approach (abbreviated as $Integration$), which comprehensively considers assertions generated by both IR-based and DL-based approaches.
Recently, hybrid approaches~\cite{sun_revisiting_2023} combining DL-based methods and IR-based methods have been proposed and proven effective. 
These combined approaches leverage the strengths of both approaches, resulting in more meaningful and useful assertions for developers.

\subsection{Test Evolution}

Hu et al.~\cite{hu_identify_2023} first use a DL-based method to generate updated test cases for the test evolution task in \textsc{Ceprot}. \textsc{Ceprot} focuses on the changes (i.e., edit sequences) between two versions of the production code and attempts to transfer these changes to the outdated test code to generate updated test code. Yaraghi et al.~\cite{yaraghi_automated_2024} propose an approach named \textsc{TaRGet}, which employs a two-step process to fine-tune a model based on the dataset. \textsc{TaRGet} focuses on repairing outdated test code. The input to the model is not limited to the related production code but also considers potential context from all changes in the project between the two versions. Recently, Liu et al.~\cite{liu_augmenting_2024} propose \textsc{Synbciatr}, which focuses on syntactic breaking changes and attempts to automatically repair outdated test cases via precise and concise Test-Repair-Oriented Contexts (TROCtx) construction.

\subsection{Large Language Models}

LLMs are large-scale models pre-trained on massive textual corpora~\cite{ouyang_training_2022, raffel_exploring_2020, vaswani_attention_2017}, demonstrating strong capabilities across various Natural Language Processing (NLP) tasks~\cite{zhao_survey_2024} and Software Engineering (SE)~\cite{zhang2023survey_se,wang_software_2024}. 
LLMs are primarily built on the Transformer~\cite{vaswani_attention_2017} architecture, which includes an encoder for input representation and a decoder for output generation.
Based on the structure, LLMs can be categorized into three types: (1) encoder-only models (e.g., CodeBERT~\cite{feng_codebert_2020}), designed for understanding tasks; (2) encoder-decoder models (e.g., CodeT5~\cite{wang_codet5_2021}), designed for translation tasks; and (3) decoder-only models (e.g., CodeGen~\cite{nijkamp_codegen_2023}), designed for generation tasks.

To improve the performance of LLMs on unseen downstream tasks, researchers often fine-tune pre-trained models using task-specific datasets~\cite{zhang2023pre,yuan2022circle}.
However, with recent advancements in LLM capabilities, particularly in in-context learning, there is a growing trend towards using prompt engineering as an alternative approach to handling downstream tasks~\cite{wang_software_2024}.
While commercial LLMs continue to dominate the top of the leaderboards recently, an increasing number of open-source models, such as CodeLlama~\cite{roziere_code_2024} and DeepSeek-Coder~\cite{guo_deepseek-coder_2024}, are emerging and demonstrating strong performance across various tasks.

\sloppy
\section{Study Design}
\label{sec:study_design} 

\subsection{Research Questions}

This study aims to answer the following research questions:

\textbf{RQ1: How does the performance of fine-tuning LLMs compare to existing approaches on three unit testing tasks?} 

\textbf{RQ2: What is the impact of various factors (including model series, model architecture, and model size) on the performance of LLMs?}

\textbf{RQ3: How do fine-tuning approaches perform compared to prompt engineering approaches?
}

\subsection{Evaluation Metrics}

Following previous studies~\cite{watson_learning_2020, yu_automated_2022, tufano_unit_2021, hu_identify_2023}, we focus on two types of metrics: (1) runtime-based metrics, which focus on the correctness and behavior of generated code within execution environments, primarily for test generation tasks; and (2) text-based metrics, which assess the syntactic and semantic accuracy of generated outputs, used for assertion generation and test evolution tasks.

\subsubsection{Runtime-based Metrics.}
We leverage five metrics to evaluate LLMs' performance on the test generation task: syntax error, compilation error, failing test, passing test, and correct test.

\begin{itemize}[leftmargin=1em]
    \item \textbf{Syntax Error.} The test has syntax errors.
    
    \item \textbf{Build Error.} The test has correct syntax but fails to build.
    
    \item \textbf{Failing Test.} The test builds but fails due to wrong assertions or expected behavior.
    
    \item \textbf{Passing Test.} The test builds and passes.
    
    \item \textbf{Correct Test.} The test passes and covers the correct focal method.
    
\end{itemize}

\subsubsection{Text-based Metrics.}

We then leverage three widely used textual metrics to evaluate the performance of LLMs on the assertion generation and test evolution tasks.

\begin{itemize}[leftmargin=1em]
    \item \textbf{Exact Match Accuracy (EM).} The EM metric~\cite{hou_large_2023}, also known as perfect accuracy, measures the proportion of correctly predicted outputs that exactly match the ground truth. A prediction is considered correct only if each token in the output sequence matches the ground truth precisely.
    
    \item \textbf{BLEU.} BLEU~\cite{papineni_bleu_2002} is an automatic metric used to evaluate the syntactic similarity between the predicted sequence and the reference sequence. It measures n-gram overlap between the translation and references, incorporating precision and brevity to ensure appropriate length.
    
    \item \textbf{CodeBLEU.} CodeBLEU~\cite{ren_codebleu_2020} is a code-aware variant of BLEU specifically designed for evaluating code synthesis tasks. Unlike BLEU, CodeBLEU incorporates syntactic similarity through Abstract Syntax Tree (AST) information and semantic similarity through data-flow analysis. This makes CodeBLEU particularly effective for assessing the quality of code generation tasks.

\end{itemize}

\subsection{Selected Models}
\label{sec:selected_models}

In our selection of LLMs, we evaluate 37 state-of-the-art models based on their architectural complexity and performance on key benchmarks. 
These models fall into three major architectural categories: encoder-only, encoder-decoder, and decoder-only. 
Our final selection includes 20 distinct model series (e.g., CodeT5, CodeGen), comprising a total of 37 models, each varying in size and architecture. 
The sizes of these models range from 60 million parameters (e.g., CodeT5-small) to 16 billion parameters (e.g., CodeGen-16b-multi), allowing us to evaluate models under different computational constraints while balancing efficiency and task performance.
We provide a detailed list of our selected models as shown in Table~\ref{tab:model_selection}.

\begin{table}[htbp]
    \centering
    \small
    \caption{Selected models in this work.}
    \label{tab:model_selection}
    \resizebox{\textwidth}{!}{
    \begin{tabular}{ll}
        \toprule
        \textbf{Architecture} & \textbf{Models} \\
        \midrule
        \textbf{Encoder-only} & CodeBERT, GraphCodeBERT, UniXcoder\\
        \midrule
        \multirow{3}{*}{\textbf{Encoder-decoder}} & CodeT5 Series (CodeT5-small, CodeT5-base, CodeT5-large) \\
        & CodeT5+ Series (CodeT5+ 220m, CodeT5+ 770m) \\
        & PLBART Series (PLBART-base, PLBART-large) \\
        \midrule
        \multirow{10}{*}{\textbf{Decoder-only}} & CodeGPT, SantaCoder, DeciCoder, CodeShell, InCoder, StarCoder \\
        & Phi Series (Phi-1, Phi-2) \\
        & CodeGen Series (CodeGen-350m-multi, CodeGen-2b-multi, CodeGen-6b-multi, CodeGen-16b-multi) \\
        & CodeGen2 Series (CodeGen2-1b-p, CodeGen2-3.7b-p) \\
        & StarCoderBase Series (StarCoderBase-1b, StarCoderBase-3b, StarCoderBase-7b, StarCoderBase-15.5b) \\
        & StarCoder2 Series (StarCoder2-3b, StarCoder2-7b, StarCoder2-15b) \\
        & CodeLlama Series (CodeLlama-7b, CodeLlama-13b) \\
        & CodeGemma Series (CodeGemma-2b, CodeGemma-7b) \\
        & DeepSeek-Coder Series (DeepSeek-Coder-1.3b-base, DeepSeek-Coder-6.7b-base) \\
        \bottomrule
    \end{tabular}
    }
\end{table}

For decoder-only LLMs, we represent tasks as auto-regressive generation tasks. 
Given an initial prompt or source sequence, the decoder model predicts the next token by sampling from the probability distribution  $P(t_i|t_1, …, t_{i-1})$, where  $t_1, ..., t_{i-1} $ represent all the tokens generated up to position $i$. This process continues iteratively until the model generates an end-of-sequence token or reaches a predefined length. 
For encoder-decoder LLMs, we represent tasks as sequence-to-sequence translation tasks. 
In these models, fine-tuning is performed on a mapped source-target pair, denoted as $mst_i = \{s_i, t_i\}$, where $s_i$ is the source sequence and $t_i$ is the target sequence. 
The fine-tuning process learns the mapping $s_i \rightarrow t_i$ as a conditional probability $P(t_i | s_i)$.
For encoder-only LLMs, we manually extend them into encoder-decoder architectures by integrating a transformer decoder stack and treat them as encoder-decoder LLMs.

After training the LLMs for each task, we apply the beam search strategy when provided with a source sequence. 
This strategy generates the predicted target sequence by selecting words from multiple candidate sequences based on the probability distribution over the vocabulary.

\subsection{Dataset}
\label{sec:dataset}

\subsubsection{Test generation task.}
In the test generation task, we utilize {Methods2Test} dataset proposed by Tufano et al.~\cite{tufano_methods2test_2022, tufano_unit_2021} to fine-tune LLMs. {Methods2Test} is a comprehensive, supervised dataset comprising test cases paired with their corresponding focal methods. This dataset is created through an extensive mining process, resulting in 780,944 pairs of JUnit tests and focal methods.

However, due to resource constraints, we are unable to use the entire dataset. To reduce the dataset and enhance its quality, we filter the dataset by following rules.
(1) \textit{Length of Tokens.} To ensure data completeness and simplicity, we filter out entries with token lengths that are either too short or too long. Specifically, we retain entries where the length of the input focal method is between 64 and 2048, and the length of the output test case is between 16 and 512.  
(2) \textit{Construction.} To normalize the construction of test cases, we only retain entries that begin with the prefix \texttt{`@Test public void'} and do not throw exceptions.
(3) \textit{Duplicated Test Cases.} We find half of the focal methods in the dataset include at least two related test cases. To increase the variety of focal methods, we randomly select one pair from each set of test cases associated with the same focal methods.
(4) \textit{Filtered repository.} To maintain the quality of the dataset, we filter out any repository with fewer than 50 pairs. Conversely, to ensure a diverse set of test cases that are not dominated by a few repositories, we randomly sample 200 pairs from any repository with more than 200 pairs.

Applying the above rules to the dataset, we obtain 56,132 pairs of data and call this dataset $Method2Test_{filter}$. We then split the filtered dataset by repositories into training, validation, and test sets using an 8:1:1 ratio for further fine-tuning.

Furthermore, previous works~\cite{tufano_unit_2021, alagarsamy_a3test_2023, ni_casmodatest_2024} primarily adopt runtime-based metrics to evaluate the performance of test generation approaches, focusing on the correct test rate.
Similar to \textsc{AthenaTest}~\cite{tufano_unit_2021}, we employ Defects4J as our benchmark dataset to evaluate LLMs' performance and use the same five evaluated projects: Apache Common Cli~\cite{2023CLI}, Apache Common Csv~\cite{2023CSV}, Google Gson~\cite{2023gson}, JFreeChart~\cite{2023Chart} and Apache Commons Lang~\cite{2023Lang}. 
These projects cover various domains, such as command-line interfaces, data processing, serialization, visualization, and utilities.

\subsubsection{Assertion generation task.}
In the assertion generation task, we use the dataset known as $Data_{old}$. This dataset is derived from a raw dataset used by \textit{ATLAS}~\cite{watson_learning_2020}. Each entry in $Data_{old}$ is referred to as a Test-Assert Pair (TAP). A TAP consists of two components: (1) a focal-test pair, which includes a test method without an assertion and its corresponding focal method; (2) assertions. 
To simplify the problem, $Data_{old}$ excludes any TAP where the assertions contain tokens that are not present in the focal-test pair. 
In total, $Data_{old}$ contains 156,760 TAPs, which are divided into training, validation, and test sets by the ratio of 8:1:1 for further fine-tuning.

\subsubsection{Test evolution task.}
In the test evolution task, we utilize the dataset proposed by Hu et al.~\cite{hu_identify_2023} Specifically for the obsolete test updating task mentioned in the paper. Each sample in the dataset consists of $<$\textit{original method, updated method, original test, updated test}$>$. The dataset contains 5,196 samples, which the authors originally split into training and test sets using a 9:1 ratio. We further create a validation set by randomly selecting 11\% of the samples from the training set, resulting in a final split ratio of 8:1:1 for the training, validation and test sets.

\subsection{Implementation Details}
\label{sec:implementation}

We implement all model training approaches using the PyTorch framework. For the studied models, we use the Hugging Face versions. For model training, we adopt the default training parameters from previous studies for each task. However, due to the variety of models and tasks, we specifically adjust some training parameters. We use the AdamW optimizer with a learning rate of 5e-5 and train for a maximum of 50 epochs for all tasks. Additionally, we apply early stopping with a patience of two epochs, meaning training halts if the loss does not decrease for two consecutive epochs.
Maximum input and output sequence lengths are adjusted according to task-specific and model-specific requirements. For the assertion generation and test evolution tasks, we set source and target sequence lengths to 512 and 256, respectively. For the test generation task, we use lengths of 2048 and 512 due to the long context information. 

We conduct our experiments on a computing cluster with several A100 nodes (8x NVIDIA A100-PCIE-40GB). We utilize DeepSpeed to accelerate training and reduce memory usage.

\sloppy
\section{Result and Analysis}
\label{sec:result_analysis}

\subsection{RQ1: Performance of LLMs on Unit Testing Tasks}
\label{sec:rq1}

\textbf{Motivation.}
Existing researches~\cite{hu_identify_2023, tufano_unit_2021, yaraghi_automated_2024} on unit testing tasks often utilize models smaller than 1 billion parameters, such as CodeBERT, CodeT5 and CodeGPT. Due to the limited size and number of models, these studies often lack comprehensive analysis of specific tasks. 
In this RQ, we aim to provide a comprehensive evaluation of LLMs' performance on three unit testing tasks. Due to the large number of LLMs, we focus our analysis on selecting representative models for each task\footnote{Complete comparison of LLMs is presented in our repository due to page limit.}. 

\subsubsection{Test Generation}
\label{sec:rq1-test_generation}

\textbf{Design.}
In the test generation task, we compare the performance of LLMs with four state-of-the-art test generation approaches from previous research: \textsc{AthenaTest}~\cite{tufano_unit_2021}, A3Test~\cite{alagarsamy_a3test_2023}, ChatUniTest~\cite{chen_chatunitest_2024}, and CasModaTest$_{~GPT3.5}$~\cite{ni_casmodatest_2024}. 
The evaluation uses five runtime-based metrics, primarily focusing on the correct test rate. Notably, due to resource constraints, we do not fine-tune several large-scale models in the test generation task.

\begin{table*}[htbp]
\centering
\small
\caption{Comparisons of LLMs with previous approaches on the test generation task using the Defects4J dataset. The best results among all LLMs are \textbf{bolded}, and the second-best results are \uline{underlined}.}
\label{tab:tg_comparisons_llm}
\begin{tabular}{lccccc}
\toprule
\textbf{Approaches/LLMs} & \textbf{Correct} & \textbf{Passing} & \textbf{Failing} & \textbf{Build Error} & \textbf{Syntax Error} \\
\midrule
\textsc{AthenaTest} & 16.21\% & 21.35\% & 26.71\% & 42.41\% & 9.49\% \\
A3Test & 40.05\% & - & - & - & - \\
ChatUniTest & 40.14\% & - & - & - & - \\
CasModaTest$_{~GPT3.5}$ & 77.16\%  & - & - & - & - \\
\midrule
CodeBERT & 5.56\% & 8.41\% & \textbf{5.99\%} & 78.90\% & 6.70\% \\
GraphCodeBERT & 11.77\% & 14.02\% & 11.19\% & 65.89\% & 8.89\% \\
UniXcoder & 6.42\% & 8.12\% & 10.48\% & 69.02\% & 12.37\% \\
CodeT5-base & 15.91\% & 18.76\% & 14.90\% & 61.36\% & \uline{4.98\%} \\
CodeT5p-220m & 17.75\% & 19.79\% & 27.07\% & 50.26\% & \textbf{2.88\%} \\
PLBART-large & 17.50\% & 20.25\% & 20.63\% & 52.96\% & 6.16\% \\
CodeGPT & 7.66\% & 9.64\% & 9.81\% & 39.67\% & 40.88\% \\
StarCoder2-7b & 20.78\% & 23.10\% & 23.16\% & 41.45\% & 12.30\% \\
CodeLlama-7b & \uline{29.50\%} & \uline{32.14\%} & \uline{7.86\%} & \uline{30.87\%} & 29.13\% \\
DeepSeek-Coder-6b & \textbf{33.68\%} & \textbf{36.02\%} & 9.00\% & \textbf{21.36\%} & 33.62\% \\
\bottomrule
\end{tabular}
\vspace{-0.5cm}
\end{table*}

\textbf{Result.}
Table~\ref{tab:tg_comparisons_llm} presents the comparison of LLMs and the four state-of-the-art approaches on the test generation task. We classify prior approaches into two categories: the first primarily leverages fine-tuning strategies for adapting models to the test generation task (e.g., \textsc{AthenaTest} and A3Test); the second employs prompt engineering based approaches to guide LLM behavior toward desired outputs without modifying model weights (e.g., ChatUniTest and CasModaTest$_{~GPT3.5}$, which predominantly uses GPT-3.5 as the underlying model).

Compared to \textsc{AthenaTest}, the encoder-only models (e.g., CodeBERT, GraphCodeBERT, and UniXcoder) underperform across correct, passing, and build error rates. Among these, GraphCodeBERT performs the best but still shows a 27.39\% reduction in the correct test rate and a 55.36\% increase in the build error rate compared to \textsc{AthenaTest}. The encoder-decoder models (e.g., CodeT5-base, CodeT5p-220m, and PLBART-large) perform similarly to \textsc{AthenaTest}, while the decoder-only models generally achieve better results. The best-performing decoder-only model, DeepSeek-Coder-6b, reaches a correct rate of 33.68\%, which is 107.77\% higher than \textsc{AthenaTest}, and has a build error rate of 21.36\%, 49.63\% lower than \textsc{AthenaTest}.

Despite their strong performance in correct test rates, decoder-only models exhibit higher syntax error rates. A manual analysis of the generated test cases marked with syntax errors reveals that these cases typically involve long contexts that exceed the models’ context window. For example, CodeGPT exhibits the highest syntax error rate, which can be attributed to its shorter context window of 1024 tokens compared to the 2048 tokens in other decoder-only models. To address this, we truncate the input context to leave enough space for the models to generate test cases. However, due to the auto-regressive fine-tuning objective, the models may attempt to complete the truncated input sequence rather than focus on generating the test case, even with adding the end-of-sequence (eos) token. In contrast, models with other architecture are less prone to this issue, likely due to the separation of tasks between the encoder and decoder components, which provides better handling of input context.

When compared to A3Test, all LLMs show inferior performance. This is primarily due to A3Test incorporating additional post-processing verification components. In addition, prompt engineering-based approaches like ChatUniTest and CasModaTest$_{~GPT3.5}$ achieve higher correct test rates. Notably, CasModaTest$_{~GPT3.5}$ reaches a 77.16\% correct rate, which is 129.10\% higher than DeepSeek-Coder-6b. These results suggest two points: first, the fine-tuning approach can be improved by a well-designed post-process. Second, with effective task segmentation, prompt construction, and a robust feedback mechanism, prompt engineering approaches can deliver exceptional performance.

\finding{1}{
Although LLMs demonstrate improvements over \textsc{AthenaTest}, they underperform compared to A3Test and the two prompt engineering-based approaches. The results suggest that fine-tuning alone is insufficient to achieve state-of-the-art performance in the test generation task, and well-designed post-processing steps are also critical.
}

\subsubsection{Assertion Generation}
\textbf{Design.}
In the assertion generation task, we compare the performance of LLMs with six state-of-the-art assertion generation approaches from prior researches~\cite{watson_learning_2020, yu_automated_2022, sun_revisiting_2023}: $ATLAS$, $IR_{ar}$, $RA_{adapt}^H$, $RA_{adapt}^{NN}$, $Integration$, and \textsc{EditAS}, using three text-based metrics. Notably, due to resource constraints, we do not fine-tune several large-scale models.

\textbf{Result.}
Table~\ref{tab:ag_comparisons_llm} compares LLMs with state-of-the-art approaches on the assertion generation task. 
We first compare the performance of the LLMs with the most relevant baseline approach, $ATLAS$, which also frames the assertion generation problem as a neural machine translation task, similar to the LLMs. 
On the $Data_{old}$ dataset, the LLMs achieve prediction EMs ranging from 50.53\% to 71.42\%, representing an improvement of 60.82\% to 127.33\% over $ATLAS$. 
Additionally, LLMs achieve the highest BLEU and CodeBLEU scores of 85.89\% and 88.25\%, respectively, which are improvements of 25.37\% and 38.76\%, over $ATLAS$. 
These significant improvements suggest that LLMs have a distinct advantage in assertion generation task when compared to the default transformer model. 

\begin{wraptable}{l}{0.5\textwidth}

\centering
\footnotesize
\caption{Comparisons of LLMs with previous approaches on the assertion generation task using the $Data_{old}$ dataset. The best results among all LLMs are \textbf{bolded}, and the second-best results are \uline{underlined}.}
\label{tab:ag_comparisons_llm}
\begin{tabular}{lccc}
\toprule
\textbf{Approaches/LLMs} & \textbf{EM} & \textbf{BLEU} & \textbf{CodeBLEU} \\
\midrule
$ATLAS$ & 31.42\% & 68.51\% & 63.60\% \\
$IR_{ar}$ & 36.26\% & 71.49\% & 71.03\% \\
$RA_{adapt}^H$ & 40.97\% & 73.28\% & 72.46\% \\
$RA_{adapt}^{NN}$ & 43.63\% & 73.95\% & 72.12\% \\
$Intergration$ & 46.54\% & 78.86\% & 73.29\% \\
\textsc{EditAS} & 53.46\% & 80.77\% & 77.00\% \\
\midrule 
CodeBERT & 54.64\% & 74.31\% & 73.89\% \\
GraphCodeBERT & 58.32\% & 74.69\% & 76.02\% \\
UniXcoder & 55.12\% & 68.92\% & 73.14\% \\
CodeT5-base & 60.26\% & \uline{85.12\%} & \uline{87.32\%} \\
CodeT5p-220m & 61.76\% & \textbf{85.89\%} & \textbf{88.25\%} \\
PLBART-base & 53.92\% & 82.09\% & 84.86\% \\
CodeGPT & 51.30\% & 61.28\% & 69.12\% \\
StarCoder2-3b & 63.29\% & 72.50\% & 77.60\% \\
CodeLlama-7b & \textbf{71.42\%} & 83.92\% & 83.34\% \\
DeepSeek-Coder-6b & \uline{70.57\%} & 82.99\% & 82.66\% \\
\bottomrule
\end{tabular}

\end{wraptable}

When compared to four retrieval-based approaches (i.e., $IR_{ar}$, $RA_{adapt}^H$, $RA_{dadpt}^{NN}$, and $intergration$), LLMs still demonstrate a clear advantage. 
We compare the best EM, BLEU, and CodeBLEU scores achieved by LLMs,  71.42\%, 85.89\%, and 88.25\%, respectively, to those achieved by the $intergration$ approach, which performs best among four approaches.
LLMs show improvements of 53.46\% in EM, 8.91\% in BLEU and 20.41\% in CodeBLEU. 
Even when compared to the most recent retrieval-enhanced learning-based approach, \textsc{EditAS}, LLMs still outperform, with improvements of 33.60\% in EM, 6.34\% in BLEU and 14.61\% in CodeBLEU. 
This further confirms the superiority of LLMs in the assertion generation task.

Additionally, we observe that while previous state-of-the-art approaches typically achieve higher BLEU scores than CodeBLEU scores, LLMs often exhibit the opposite trend, producing higher CodeBLEU scores compared to BLEU. 
This discrepancy may be due to the extensive code-related corpus used during the LLMs’ pre-training phase, which enhances their understanding of code structure and boosts their performance on code-related tasks.

\finding{2}{
Compared to existing state-of-the-art approaches, LLMs demonstrate significant improvements across all three text-based metrics. 
Notably, LLMs tend to achieve higher CodeBLEU scores than BLEU scores, which is opposite of the trend observed in previous approaches, suggesting that LLMs have a better understanding of code structure.
}

\subsubsection{Test Evolution}
\label{sec:rq1-test-evolution}

\textbf{Design.}
In the test evolution task, we compare the performance of LLMs against the \textsc{Ceprot}~\cite{hu_identify_2023} approach and two traditional machine learning methods, KNN and NMT, using three text-based metrics.

\textbf{Result.}
Table~\ref{tab:tu_comparisons_llm} compares LLMs and three approaches on the test evolution task, showing that LLMs consistently outperform the two machine learning methods, KNN and NMT. 
The prediction EM for LLMs ranges from 6.15\% to 35.58\%, showing a substantial improvement of 23.0\% to 611.6\% over NMT. 
For the CodeBLEU metric, LLMs achieve scores ranging from 46.42\% to 86.35\%, marking an improvement of 23.46\% to 129.65\% over KNN. These results highlight that LLM-based methods significantly outperform traditional machine learning approaches in the test evolution task.

\begin{wraptable}{l}{0.5\textwidth}
\centering
\footnotesize
\caption{Comparisons of LLMs with previous approaches on the test evolution task. The best results among all LLMs are \textbf{bolded}, and the second-best results are \uline{underlined}.}
\label{tab:tu_comparisons_llm}
\begin{tabular}{lccc}
\toprule
\textbf{Apporaches/LLMs} & \textbf{EM} & \textbf{BLEU} & \textbf{CodeBLEU}  \ \\
\midrule
KNN & 3.9\% & - & 37.6\% \\
NMT & 5.0\% & - & 32.3\% \\ 
\textsc{Ceprot} & 12.3\% & - & 63.1\% \\
\midrule
CodeBERT & 6.15\% & 37.72\% & 46.42\% \\
GraphCodeBERT & 12.69\% & 63.49\% & 67.37\% \\
UniXcoder & 10.58\% & 58.03\% & 63.12\% \\
CodeT5-base & 12.88\% & 78.31\% & 78.31\% \\
CodeT5p-220m & 17.12\% & 81.54\% & 81.54\% \\
PLBART-base & 12.88\% & 77.88\% & 78.45\% \\
CodeGPT & 15.38\% & 79.77\% & 81.58\% \\
StarCoder2-15b & 29.04\% & 83.45\% & 84.71\% \\
CodeLlama-13b & \uline{34.62\%} & \textbf{85.20\%} & \uline{86.13\%} \\
DeepSeek-Coder-6b & \textbf{35.58\%} & \uline{84.48\%} & \textbf{86.35\%} \\

\bottomrule
\end{tabular}
\end{wraptable}

We then compare the performance of LLMs with the \textsc{Ceprot} approach. Interestingly, we observe a similar situation as in the test generation task. 
The encoder-only models perform the worst, while encoder-decoder models achieve results similar to \textsc{Ceprot}, and decoder-only models generally perform better. 
The top-performing model, DeepSeek-Coder-6b, achieves 35.58\% EM and 86.35\% CodeBLEU, representing improvements of 189.27\% and 36.85\%, respectively. 
Furthermore, as described in Section~\ref{sec:implementation}, we use a tuple containing the original method, the updated method, and the original test as model inputs, thereby removing the edit sequence compared to \textsc{Ceprot}. 
We find that the edit sequence is too long to be fully incorporated into the input vector, and truncating it typically results in the loss of valuable information, leading us to remove the edit sequence entirely.
Since \textsc{Ceprot} also utilizes CodeT5 as its base model, we compare the performance of the standalone CodeT5-base model directly with \textsc{Ceprot}.
The standalone CodeT5-base model achieves 12.88\% in EM and 78.31\% in CodeBLEU, showing an improvement of 4.7\% in EM and 24.1\% in CodeBLEU compared to \textsc{Ceprot}. This indicates that reallocating space from the edit sequence to other content improves the LLMs’ performance, particularly in terms of the CodeBLEU score.

\finding{3}{
Overall, all LLMs outperform traditional ML-based methods, and most LLMs also surpass the performance of \textsc{Ceprot}. Furthermore, comparing CodeT5-base and \textsc{CEPROT} suggests that reallocating space from the edit sequence to other content can improve performance.
}

\subsection{RQ2: Impact of Various Factors on LLMs' Performance}
\label{sec:rq2}

\textbf{Motivation.}
In addition to evaluating specific LLMs, we are also interested in identifying the key factors of high-performing LLMs. 
Our analysis is based on three factors: model series, model architecture, and model size. 
In this RQ, we aim to determine how each of these factors influences the performance of LLMs.
The model series groups models within the same series into one category, as shown in Section~\ref{sec:selected_models}. For instance, the CodeT5 series includes CodeT5-small, CodeT5-base, and CodeT5-large. Model architecture refers to three types: encoder-only, encoder-decoder, and decoder-only. Model size pertains to the number of parameters in each model, which in our study ranges from 60 million to 16 billion.

\subsubsection{The Impact of Model Series}

\textbf{Design.}
In this section, we explore the performance of different model series by analyzing 17 different representative model series across three unit testing tasks.

\begin{table}[htbp]
    \centering
    \small
    \caption{Performane of different series LLMs across three tasks. The best results among all LLMs are highlighted in \textbf{bold}, and the second-best results are \uline{underlined}.}
    \label{tab:rq2-all_result}
    \resizebox{\textwidth}{!}{
    \begin{tabular}{cccccccccccc}
    \toprule
\multirow{2}{*}{\textbf{LLMs}} & \multicolumn{5}{c}{\textbf{Test Generation}} & \multicolumn{3}{c}{\textbf{Assertion Generation}} & \multicolumn{3}{c}{\textbf{Test Evolution}} \\
\cmidrule(lr){2-6} \cmidrule(lr){7-9} \cmidrule(lr){10-12}
& \textbf{Correct} & \textbf{Passing} & \textbf{Failing} & \textbf{Build Error} & \textbf{Syntax Error} & \textbf{EM} & \textbf{BLEU} & \textbf{CodeBLEU} & \textbf{EM} & \textbf{BLEU} & \textbf{CodeBLEU}\\
    \midrule
    CodeBERT & 5.56\% & 8.41\% & \textbf{5.99\%} & 78.90\% & 6.70\% & 54.64\% & 74.31\% & 73.89\% & 6.15\% & 37.72\% & 46.42\%  \\ 
    GraphCodeBERT & 11.77\% & 14.02\% & 11.19\% & 65.89\% & 8.89\% & 58.32\% & 74.69\% & 76.02\% & 12.69\% & 63.49\% & 67.37\%  \\ 
    UniXcoder & 6.42\% & 8.12\% & 10.48\% & 69.02\% & 12.37\% & 55.12\% & 68.92\% & 73.14\% & 10.58\% & 58.03\% & 63.12\%  \\ 
    \midrule
    CodeT5 & 16.63\% & 18.95\% & 20.02\% & 57.30\% & \textbf{3.73\%} & 58.57\% & \uline{84.22\%} & \uline{86.67\%} & 12.76\% & 76.55\% & 76.75\%  \\ 
    CodeT5p & 15.94\% & 18.64\% & 24.69\% & 51.25\% & \uline{5.44\%} & 61.02\% & \textbf{85.47\%} & \textbf{87.67\%} & 19.52\% & 79.31\% & 80.43\%  \\ 
    PLBART & 14.96\% & 16.96\% & 16.82\% & 59.51\% & 6.71\% & 54.51\% & 82.31\% & 85.01\% & 14.04\% & 77.84\% & 78.56\%  \\ 
    \midrule
    CodeGPT & 7.66\% & 9.64\% & 9.81\% & 39.67\% & 40.88\% & 51.30\% & 61.28\% & 69.12\% & 15.38\% & 79.77\% & 81.58\%  \\ 
    InCoder & 17.50\% & 19.32\% & 21.56\% & 50.73\% & 8.39\% & 62.24\% & 77.17\% & 75.50\% & 26.15\% & 79.95\% & 83.41\%  \\ 
    CodeGen & 16.80\% & 19.24\% & 11.92\% & 37.24\% & 31.60\% & 62.53\% & 75.59\% & 77.05\% & 27.93\% & 81.97\% & 83.17\%  \\ 
    CodeGen2 & - & - & - & - & - & 62.74\% & 74.40\% & 77.02\% & \uline{28.56\%} & 83.31\% & 84.72\%  \\ 
    StarCodeBase & \uline{27.82\%} & \uline{31.02\%} & 14.65\% & 34.69\% & 19.64\% & 65.24\% & 76.50\% & 79.35\% & \uline{28.56\%} & 81.74\% & 83.82\%  \\ 
    StarCode2 & 21.22\% & 23.74\% & 21.10\% & 42.15\% & 13.02\% & 64.03\% & 75.40\% & 78.16\% & 24.55\% & 81.16\% & 83.16\%  \\ 
    Phi & 14.77\% & 19.51\% & 12.04\% & 53.85\% & 14.61\% & 55.37\% & 67.78\% & 72.34\% & 18.56\% & 79.79\% & 80.29\%  \\ 
    CodeLlama & \textbf{29.50\%} & \textbf{32.14\%} & 7.86\% & \uline{30.87\%} & 29.13\% & \textbf{71.42\%} & 83.92\% & 83.34\% & \textbf{34.62\%} & 83.21\% & \uline{85.33\%}  \\ 
    CodeGemma & 22.84\% & 25.50\% & 15.03\% & 50.37\% & 9.10\% & 61.06\% & 74.26\% & 77.79\% & 27.88\% & \textbf{84.34\%} & 85.06\%  \\ 
    DeepSeekCoder & 25.63\% & 29.20\% & 11.03\% & \textbf{28.78\%} & 31.00\% & \uline{68.10\%} & 81.46\% & 81.23\% & \textbf{34.62\%} & \uline{84.19\%} & \textbf{86.05\%}  \\ 
    SantaCoder & 11.55\% & 15.44\% & \uline{7.38\%} & 39.33\% & 37.85\% & 57.60\% & 73.44\% & 74.66\% & 27.12\% & 82.46\% & 83.85\% \\ 
    \bottomrule
    \end{tabular}
    }
\end{table}

\textbf{Result.}
Table~\ref{tab:rq2-all_result} demonstrates the performance of different series of LLMs across three unit testing tasks. We calculate the average score of a LLM series to represent the performance of this LLM series. According to Table~\ref{tab:rq2-all_result}, CodeLlama leads in performance, achieving the highest EM scores in both the assertion generation (71.42\%) and test evolution (34.62\%) tasks, as well as the highest correct test and passing rates in the test generation task, 29.50\% and 32.14\%, respectively. DeepSeek-Coder and StarCodeBase also perform well: DeepSeek-Coder ranks second in EM for assertion generation (68.10\%) and tops the test evolution task in EM and CodeBLEU scores (34.62\%). StarCodeBase achieves strong results in test generation, with the second-highest correct test rate at 27.82\% and the passing rate at 31.02\%, as well as a high EM score at 28.56\% in test evolution. CodeT5p and CodeT5 series models excel in BLEU and CodeBLEU scores in assertion generation.

On the other hand, CodeBERT performs the worst across tasks, with the lowest correct test rate at 5.56\%, the passing rate at 8.41\%, and the highest build error rate at 78.90\% in the test generation task. It also has the lowest EM, BLEU, and CodeBLEU scores in the test evolution task, 6.15\%, 37.72\%, and 46.42\%, respectively. Similarly, CodeGPT has the lowest EM, BLEU, and CodeBLEU scores for assertion generation, 51.30\%, 61.28\%, and 69.12\%, respectively, and the highest syntax error rate in the test generation task.

\finding{4}{
CodeLlama outperforms other models across all tasks, with the highest EM, correct test rate, and passing rate. DeepSeekCoder and StarCodeBase also show strong performance, particularly in assertion generation and test generation tasks. In contrast, CodeBERT and CodeGPT consistently perform the worst, with low EM and high error rates.
}

\subsubsection{The Impact of Model Architecture}

\begin{figure}[htbp]
    \centering
    \includegraphics[width=\textwidth]{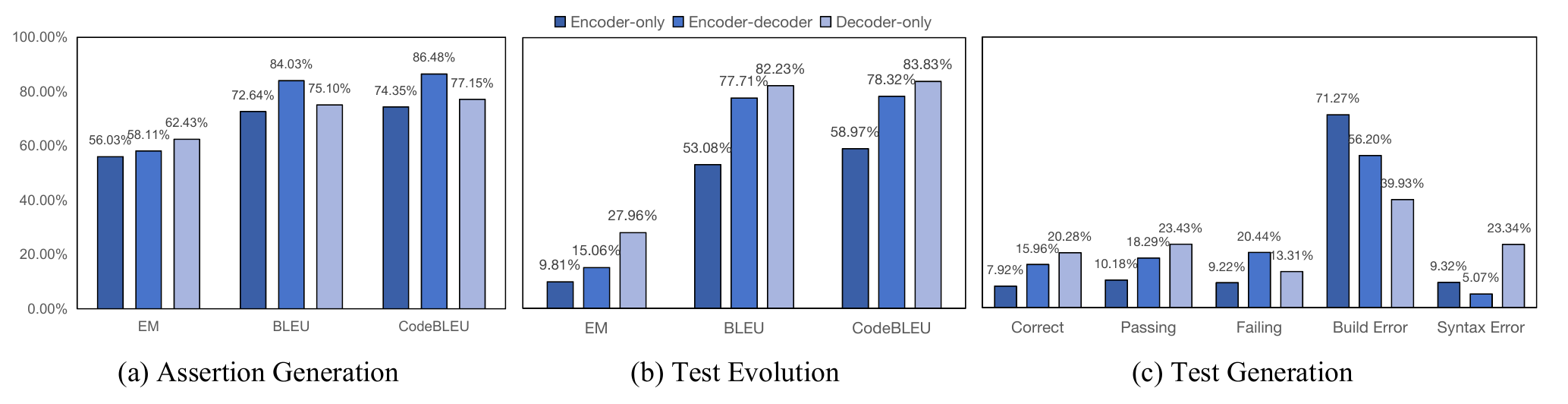}
    \caption{Performance of LLMs with different architectures across three tasks.}
    \label{fig:rq2_1_all_tasks}
\end{figure}

\textbf{Design.}
In this section, we examine how model architecture influences performance. To provide a clearer comparison across architectures, we calculate the average metrics for each architecture.

\textbf{Result.}
Fig.~\ref{fig:rq2_1_all_tasks} shows the performance of LLMs with different architectures across three tasks. 
In the assertion generation task, decoder-only models achieve the highest average EM score at 62.43\%, followed by encoder-decoder models at 58.11\% and encoder-only models at 56.03\%. However, when considering BLEU and CodeBLEU scores, encoder-decoder models outperform others with an average BLEU of 84.03\% and CodeBLEU of 86.48\%. Decoder-only models perform less well, with BLEU at 72.64\% and CodeBLEU at 74.35\%.
In the test evolution task, the EM scores follow a similar pattern as assertion generation, with decoder-only models leading at 27.96\%. Unlike assertion generation, decoder-only models also outperform BLEU and CodeBLEU, with averages of 82.23\% and 83.83\%, respectively. Encoder-decoder models lag behind with lower EM (15.05\%), BLEU (77.71\%), and CodeBLEU (78.32\%), while encoder-only models perform the worst across all metrics.
In the test generation task, decoder-only models achieve the highest correct test rate (20.28\%) and passing rate (23.43\%), alongside the lowest build error rate (39.93\%). This suggests that decoder-only models generate more correct test cases than the other architectures. Encoder-decoder models perform moderately, while encoder-only models perform the worst. However, when it comes to syntax error rates, decoder-only models exhibit the highest error rate at 23.34\%, significantly higher than the other two architectures. We speculate that this is due to the auto-regressive objective of decoder-only models analyzed in Section~\ref{sec:rq1-test_generation}, which may lead them to complete truncated input sequences rather than focus on generating the test case, resulting in higher syntax errors.

\finding{5}{
Decoder-only models generally perform best across all three tasks, particularly in EM and correctness metrics, but they exhibit higher syntax error rates than the other two architecture models. Encoder-decoder models excel in BLEU and CodeBLEU scores on the assertion generation task, while encoder-only models consistently underperform across all tasks.
}

\subsubsection{The Impact of Model Size}

\textbf{Design.}
In this section, we explore how model size influences performance across three tasks. We collect the parameter sizes of LLMs and list them on the scatter plot for further analysis.

\begin{figure}[htbp]
    \centering
    \includegraphics[width=\textwidth]{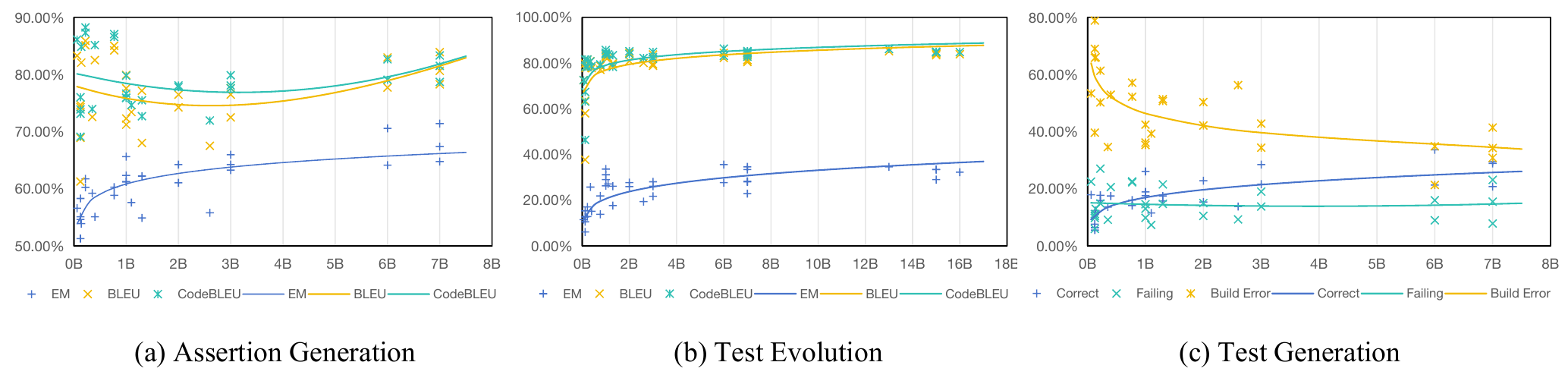}
    \caption{Performance of LLMs with different sizes across three tasks.}
    \label{fig:rq2_3_all_tasks}
\end{figure}

\textbf{Result.}
Fig.~\ref{fig:rq2_3_all_tasks} presents scatter plots of model performance in relation to parameter size, with trend lines added for clarity. In the assertion generation task, the EM score increases as the model size grows, following a pattern of rapid improvement initially, followed by slower gains. BLEU and CodeBLEU scores show a different pattern, first decreasing and then increasing as the size grows, influenced by the strong performance of CodeT5 and CodeT5p models.
In the test evolution task, all three metrics—EM, BLEU, and CodeBLEU—improve with increasing model size, again showing rapid early improvement that slows as size increases, indicating a clear positive correlation between size and performance.
In the test generation task, the correct test rate increases with model size while the build error rate decreases, suggesting that larger models tend to produce more accurate test cases with fewer build errors. However, the failing rate does not significantly decrease, indicating that even with larger models, many generated test cases still fail to pass.

\finding{6}{
Large-scale models generally perform better across all tasks, with improved metrics like EM, correct test rate, and reduced build error rates. However, even as model size increases, the failing rate remains high, suggesting that test cases often fail regardless of model size.
}

\subsection{RQ3: Comparison between the Fine-tuning and Prompt Engineering Approaches}
\label{sec:rq3}

\textbf{Motivation.}
Recently, with the development of LLMs such as ChatGPT, there is a growing trend towards using prompt engineering as an alternative approach to handling downstream tasks~\cite{wang_software_2024}. Unlike fine-tuning LLMs with specific downstream datasets, prompt engineering focuses on designing advanced prompts that incorporate task-specific information, aiming to leverage the existing knowledge within the LLM to solve the problem. 

\textbf{Design.}
In this RQ, we aim to evaluate and compare the effectiveness of fine-tuning and prompt engineering approaches across three unit testing tasks. We examine six widely used instruction-based LLMs (open and closed-sourced), including GPT-3.5-Turbo, Llama3.1-8b, Llama3.1-70b, Llama3-8b, Qwen2-7b, and Qwen2-72b. For open-source models, we select the instruction-tuned versions to support prompt-based testing. 
Additionally, We apply the same input data across models and adopt a zero-shot learning setup for prompt construction\footnote{Detailed prompt template is presented in our repository due to page limit.}.
Considering that instruction-based LLMs are not explicitly trained for specific downstream tasks, their generated outputs may deviate in format from the ground truth. To address this, we implement two strategies. First, we design structured prompts to guide the model toward desired output formats. Second, we apply post-processing steps to further refine the output.
For tasks like assertion generation and test evolution, which are evaluated using text-based metrics, we extract the code output from the model’s response as the predicted output. Then, we use a BERT tokenizer to tokenize both the predicted output and the ground truth, followed by detokenization to standardize formatting. This process unifies delimiters such as whitespace and line breaks and removes special code structures, allowing us to focus purely on semantic content.
For test generation tasks, which are evaluated using runtime-based metrics, we also extract the code output from the model response but limit it to the first unit test case. Since the fine-tuning approach generates one test case per focal method, this selective extraction ensures a fair comparison by aligning the number of generated test cases between the two approaches.

\begin{table}[!ht]
    \centering
    \small
    \caption{Comparison between fine-tuning and prompt engineering approaches across three tasks. The best results among all LLMs are \textbf{bolded}, and the second-best results are \uline{underlined}.}
    \label{tab:rq3-result}
    \resizebox{\textwidth}{!}{
    \begin{tabular}{cccccccccccc}
    \toprule
\multirow{2}{*}{\textbf{LLMs}} & \multicolumn{5}{c}{\textbf{Test Generation}} & \multicolumn{3}{c}{\textbf{Assertion Generation}} & \multicolumn{3}{c}{\textbf{Test Evolution}} \\
\cmidrule(lr){2-6} \cmidrule(lr){7-9} \cmidrule(lr){10-12}
& \textbf{Correct} & \textbf{Passing} & \textbf{Failing} & \textbf{Build Error} & \textbf{Syntax Error} & \textbf{EM} & \textbf{BLEU} & \textbf{CodeBLEU} & \textbf{EM} & \textbf{BLEU} & \textbf{CodeBLEU}\\
    \midrule
    CodeT5p-220m & 17.75\% & 19.79\% & 27.07\% & 50.26\% & 2.88\% & 61.76\% & \textbf{85.89\%} & \textbf{88.25\%} & 17.12\% & 81.54\% & 81.54\%  \\ 
    CodeLlama-7b & 29.50\% & 32.14\% & \textbf{7.86\%} & \uline{30.87\%} & 29.13\% & \textbf{71.42\%} & \uline{83.92\%} & \uline{83.34\%} & \uline{34.62\%} & 81.22\% & \uline{84.52\%}  \\ 
    DeepSeek-Coder-6b & 33.68\% & 36.02\% & 9.00\% & \textbf{21.36\%} & 33.62\% & \uline{70.57\%} & 82.99\% & 82.66\% & \textbf{35.58\%} & \uline{84.48\%} & \textbf{86.35\%}  \\ 
    \midrule
    GPT-3.5 & \textbf{49.16\%} & \textbf{51.03\%} & 11.33\% & 34.44\% & 3.20\% & 2.97\% & 32.41\% & 25.73\% & 15.96\% & \textbf{86.20\%} & 83.61\%  \\ 
    Llama3.1-8b & 30.34\% & 35.04\% & 10.22\% & 53.99\% & 0.75\% & 0.22\% & 20.93\% & 19.46\% & 4.23\% & 59.64\% & 65.25\%  \\ 
    Llama3.1-70b & 31.02\% & 33.38\% & \uline{7.99\%} & 58.09\% & \textbf{0.54\%} & 3.89\% & 31.58\% & 27.46\% & 5.39\% & 58.40\% & 63.37\%  \\ 
    Llama3-8b & \uline{36.88\%} & \uline{38.73\%} & 16.98\% & 43.60\% & \uline{0.69\%} & 0.14\% & 11.49\% & 18.10\% & 5.96\% & 71.16\% & 76.29\%  \\ 
    Qwen2-7b & 16.54\% & 21.39\% & 18.93\% & 57.51\% & 2.17\% & 0.91\% & 10.17\% & 22.49\% & 0.19\% & 42.70\% & 51.07\%  \\ 
    Qwen2-72b & 29.48\% & 31.13\% & 21.81\% & 45.60\% & 1.46\% & 4.89\% & 27.31\% & 27.46\% & 3.46\% & 54.67\% & 62.18\%  \\ 
    \bottomrule
    \end{tabular}
    }
\end{table}

\textbf{Result.}
Table~\ref{tab:rq3-result} presents a performance comparison between fine-tuned models and instruction-based models. For clarity, we focus on three representative fine-tuned models: CodeT5p-220m, CodeLlama-7b, and DeepSeek-Coder-6b.
In the test generation task, GPT-3.5 achieves a correct rate of 49.16\% and a passing rate of 51.03\%, surpassing the best-fine-tuned model, DeepSeek-Coder-6b, which achieves 45.96\% and 41.67\%, respectively. This suggests promising potential for instruction-based LLMs in test generation tasks.
In the assertion generation task, however, instruction-based models underperform, with an EM score of only 4.89\%, significantly lower than that of fine-tuned models. BLEU and CodeBLEU scores also lag; even the best-performing instruction model, GPT-3.5, reaches only 32.41\% on BLEU and 25.73\% on CodeBLEU. We manually analyze the predicted outputs from instruction-based models and observe a tendency to omit fully qualified names, likely because the model assumes that all necessary imports are included.  For instance, the prediction might generate \texttt{`assertNull(test)'}, while the ground truth specifies \texttt{`org.junit.Assert.assertNull(test)'}, contributing to lower performance scores.
In the test evolution task, instruction-based models show a wide performance range, with EM scores spanning 0.19\% to 15.96\%, BLEU from 42.70\% to 86.20\%, and CodeBLEU from 51.07\% to 83.61\%. Although instruction-based models perform poorly on EM compared to fine-tuned models, they demonstrate promising BLEU and CodeBLEU scores. Notably, GPT-3.5 achieves the highest BLEU score at 86.20\% and ranks top in CodeBLEU at 83.61\%.

Additionally, all the instruction-based models perform a low syntax error rate range from 0.54\% to 3.2\%, showing a promising understanding of code structure. And GPT-3.5 consistently achieves the best overall results, ranking first in seven metrics among the six models. 
The results of the prompt engineering approach demonstrate its potential in unit test-related tasks. Notably, this study employs only zero-shot learning to construct prompts and generate model responses. Recently, however, advanced in-context learning techniques, such as few-shot learning, Chain of Thought (CoT), and Retrieval Augmented Generation (RAG), have emerged, enhancing LLMs’ capacity to produce accurate responses. Leveraging these advanced techniques could enable the prompt engineering approach to perform even better in common unit testing tasks.

\finding{7}{
In summary, the prompt engineering approach shows promising results compared to the fine-tuning approach, particularly in test generation and test evolution tasks, highlighting its potential in unit test-related tasks. Leveraging advanced in-context learning methods, like few-shot learning, CoT, and RAG, could further enhance its performance.
}

\sloppy
\section{Discussion}
\label{sec:discussion}

\subsection{Discussion of Potential Data Leakage}
\label{sec:discussion_utbench}

The issue of data leakage, as discussed by Zhang et al.~\cite{zhang2023critical}, poses a critical risk to the evaluation of LLMs in the SE community.
In this section, we concentrate on the test generation task, for which we employ Defects4J as an evaluation dataset. 
To mitigate potential data leakage risk, we test LLMs' performance on a real-world industrial dataset\footnote{Due to the confidential policy of the company, we hide the information of internal subjects.}.
Unlike Defects4J, which partially relies on JUnit 3, the internal dataset adopts a unified Maven architecture with the JUnit 4 testing framework. This setup aligns with mainstream development practices and provides a more accurate measure of LLMs' capability to generate test cases for contemporary Java projects.

\begin{table*}[htbp]
\centering
\small
\caption{Performance of different LLMs on the internal dataset. The best results among all LLMs are \textbf{bolded}, and the second-best results are \uline{underlined}.}
\label{tab:discussion_utbench}
\begin{tabular}{lccccc}
\toprule
\textbf{LLMs} & \textbf{Correct} & \textbf{Passing} & \textbf{Failing} & \textbf{Build Error} & \textbf{Syntax Error} \\
\midrule
UniXcoder         & 5.39\%  & 6.59\%  & \textbf{1.95\%}  & 74.10\%  & 17.37\% \\
CodeT5-large      & 14.07\% & 16.32\% & 10.03\% & 66.77\%  & 6.89\%  \\
CodeT5p-770m      & 18.56\% & 19.61\% & 5.54\%  & 71.11\%  & \uline{3.74\%}  \\
DeepSeek-Coder-6b & 31.14\% & 31.59\% & 2.84\%  & \textbf{29.64\%}  & 35.93\% \\
CodeLlama-7b      & 25.00\% & 25.45\% & \uline{2.10\%}  & 34.28\%  & 38.17\% \\  
StarCoder2-7b     & 23.05\% & 24.40\% & 6.59\%  & 42.66\%  & 26.35\% \\
StarCoderBase-7b  & 23.65\% & 27.10\% & 2.40\%  & 42.96\%  & 27.54\% \\
CodeGen-6b        & 23.50\% & 23.95\% & \uline{2.10\%}  & 35.78\%  & 38.17\% \\
\midrule
GPT-3.5           & \textbf{52.54\%} & \textbf{54.34\%} & 12.43\% & \uline{32.04\%}  & \textbf{1.20\%}  \\
Llama3.1-8b       & 21.71\% & 22.01\% & 17.96\% & 51.50\%  & 8.53\% \\
Llama3.1-70b      & \uline{51.35\%} & \uline{52.40\%} & 10.48\% & {32.63\%}  & 4.49\% \\
Llama3-8b         & 28.44\% & 29.94\% & 10.33\% & 54.34\%  & 5.39\%  \\
Qwen2-7b          & 32.34\% & 32.49\% & 10.03\% & {50.90\%}  & 6.59\% \\
Qwen2-72b         & 42.37\% & 44.61\% & 7.63\% & 42.37\%  & 5.39\%  \\
\bottomrule
\end{tabular}
\end{table*}

Table~\ref{tab:discussion_utbench} presents the performance of LLMs on the internal dataset. For simplicity, we select several representative models from both fine-tuned and instruction-based models that perform well on Defects4J dataset. 
Among fine-tuned models, DeepSeek-Coder-6b achieves the highest performance with a correct rate of 31.14\% and the lowest build error rate of 29.64\%. The encoder-only model UniXcoder performs the lowest, with a correct rate of only 5.39\%, while encoder-decoder models like CodeT5-large and CodeT5p-770m have the lowest syntax error rate at 6.89\% and 3.74\%, respectively. Overall, results on the internal dataset for fine-tuned models closely align with their performance on Defects4J.
In instruction-based models, GPT-3.5 achieves the best results, ranking highest in correct rate, passing rate, and syntax error rate, at 52.54\%, 54.34\%, and 1.20\%, respectively. Llama3.1-70b and Qwen2-72b generally perform well, occupying second and third places across metrics, with Llama3.1-70b achieving a correct rate of 51.35\% and Qwen2-72b at 42.37\%.  

Instruction-based models generally achieve higher correct rates and lower syntax error rates than fine-tuned models, especially when the model size exceeds 70 billion parameters, showing trends similar to Defects4J dataset. 
For clarity, we analyze three model groups: (1) fine-tuned models around 7 billion parameters (e.g., DeepSeek-Coder-6b, CodeLlama-7b, StarCoder2-7b, StarCoderBase-7b, CodeGen-6b), (2) instruction-based models around 7 billion parameters (e.g., Llama3.1-8b, Llama3-8b, Qwen2-7b), and (3) instruction-based models exceeding 70 billion parameters (e.g., GPT-3.5, Llama3.1-70b, Qwen2-72b).
First, the 7-billion-parameter fine-tuned models achieve correct rates between 23.05\% and 31.14\%, averaging 25.27\%. 
Second, the 7-billion-parameter instruction-based models show a modest improvement, with correct rates between 21.71\% and 32.34\% and an average of 27.50\%, marking an 8.82\% increase. 
Third, instruction-based models with over 70 billion parameters achieve correct rates between 42.37\% and 52.54\%, with an average of 48.75\%, reflecting substantial improvements of 92.92\% over the 7-billion fine-tuned models and 77.27\% over the 7-billion instruction-based models.

Overall, these results highlight two main insights. First, among models with similar parameter counts, the performance gap between fine-tuned and instruction-based models is relatively modest. Second, as instruction-based models grow in size and capability, the prompt engineering approach shows strong potential to achieve even higher performance levels.

\summary{
While large-scale instruction-based models outperform fine-tuned models overall, instruction-based models with the same parameters scale show slight gains over fine-tuned counterparts. Instruction-based models show promise as they scale in size and capability.
}

\subsection{Discussion of Bug Detection Capability}
\label{sec:discussion_fault}
Evaluating the bug detection capability of generated test cases is essential for understanding their practical effectiveness in real-world development. This section evaluates the ability of generated test cases to detect bugs within a program.
To do this, we select several representative models and run their generated test cases on both buggy and fixed versions of five projects aforementioned. Following the previous works~\cite{dinella_toga_2022, hossain_togll_2024, liu_more_2023a, hossain_neuralbased_2023a}, the generated tests can be divided into four groups based on the execution results: True Positive (TP), True Negative (TN), False Positive (FP), and False Negative (FN). Here, Positive indicates that a test case fails on the buggy program, while Negative means it passes on the buggy version. True or False further specifies whether the test case correctly or incorrectly passes/fails on the fixed version.
A test is considered as a bug-finding test if it fails on the buggy program but passes on the fixed version.
We assess each model’s effectiveness and usability by calculating both the number of bugs found and the precision of the test cases. Precision is defined as $\#TP/(\#FP + \#TP)$, representing the proportion of generated test cases that successfully identify bugs among all failing test cases. A higher precision indicates fewer irrelevant test cases, reducing the validation effort required from developers.

\begin{table}[ht]
\centering
\small
\caption{The performance of LLMs for bug-detection on Defects4J}
\label{tab:bug-detection}
\begin{tabular}{lcccc}
\toprule
\textbf{Model} & \textbf{BugFound} & \textbf{Precision} & \textbf{\#TP} & \textbf{\#FP} \\
\midrule
CodeBERT & 0 & 0.00\% & 0 & 4535 \\
GraphCodeBERT & 1 & 0.02\% & 1 & 4118 \\
UniXcoder & 0 & 0.00\% & 0 & 4247 \\
CodeT5-base & 4 & 0.16\% & 7 & 4262 \\
CodeT5p-220m & 1 & 0.02\% & 1 & 4131 \\
PLBART-large & 1 & 0.03\% & 1 & 3931 \\
CodeGPT & 1 & 0.04\% & 1 & 2643 \\
StarCoder2-3b & 3 & 0.21\% & 7 & 3306 \\
CodeLlama-7b & 3 & 0.43\% & 9 & 2069 \\
DeepSeek-Coder-6b & 8 & 0.74\% & 12 & 1622 \\
\bottomrule
\end{tabular}
\end{table}

Table~\ref{tab:bug-detection} presents the bug detection performance of generated test cases on Defects4J. Overall, most models perform poorly, typically identifying zero or only one bug. Precision rates are similarly low, ranging from 0.00\% to 0.74\%, indicating that most LLM-generated test cases fail to detect bugs. The best-performing model, DeepSeek-Coder-6b, identifies 8 out of 163 bugs across five projects, achieving a precision of 0.74\%. This means that, even in the best case, a developer must review approximately 135 failing tests to find a single bug.
Further analysis of FP samples reveals that build errors account for 69.24\% to 92.94\% of all failing test cases, which is the primary reason for these failures. These findings suggest that simply generating test cases through fine-tuning alone is insufficient for effective bug detection. Improving the bug detection capability of generated test cases remains a critical challenge for future work.

\summary{
LLM-generated test cases exhibit limited bug detection effectiveness, with low precision and high rates of build errors among failing tests, highlighting the need for improved methodologies to enhance bug detection capabilities.
}

\subsection{Discussion of Metrics Comparison}
\label{sec:metrics_comparison}

In addition to runtime-based metrics, we incorporate text-based metrics to assess LLMs' performance in the test generation task on the $Methods2Test_{filter}$ dataset. Notably, we include a syntax error rate, a measure that provides valuable insights without requiring code execution.

\begin{table}[htbp]
    \centering
    \small
    \caption{Comparison of LLMs across different metrics on Defects4J and $Methods2Test_{filter}$ benchmarks.}
    \label{tab:metrics_compirason}
    \resizebox{\textwidth}{!}{
    \begin{tabular}{lcccccccccc}
    \toprule
    \multirow{2}{*}{\textbf{LLMs}}
    & \multicolumn{5}{c}{\textbf{Defects4J}} & \multicolumn{4}{c}{\textbf{Methods2Test$_{\textbf{filter}}$}} \\
    \cmidrule(lr){2-6} \cmidrule(lr){7-10}
     & \textbf{Correct} & \textbf{Passing} & \textbf{Failing} & \textbf{Build Error} & \textbf{Syntax Error} & \textbf{EM} & \textbf{BLEU} & \textbf{CodeBLEU} & \textbf{Syntax Error} \\

    \midrule
    CodeBERT & 5.56\% & 8.41\% & 5.99\% & 78.90\% & 6.70\% & 0.21\% & 11.96\% & 18.23\% & 9.57\% \\
    GraphCodeBERT & 11.77\% & 14.02\% & 11.19\% & 65.89\% & 8.89\% & 0.55\% & 12.65\% & 18.51\% & 10.99\% \\
    UniXcoder & 6.42\% & 8.12\% & 10.48\% & 69.02\% & 12.37\% & 0.46\% & 12.41\% & 18.82\% & 10.23\% \\
    CodeT5-base & 15.91\% & 18.76\% & 14.90\% & 61.36\% & 4.98\% & 0.29\% & 6.11\% & 23.89\% & 10.09\% \\
    CodeT5p-220m & 17.75\% & 19.79\% & 27.07\% & 50.26\% & 2.88\% & 0.30\% & 12.11\% & 23.00\% & 3.73\% \\
    PLBART-large & 17.50\% & 20.25\% & 20.63\% & 52.96\% & 6.16\% & 0.05\% & 9.43\% & 22.16\% & 6.47\% \\
    CodeGPT & 7.66\% & 9.64\% & 9.81\% & 39.67\% & 40.88\% & 0.09\% & 8.28\% & 21.50\% & 15.47\% \\
    StarCoder2-7b & 20.78\% & 23.10\% & 23.16\% & 41.45\% & 12.30\% & 0.27\% & 5.56\% & 22.34\% & 5.92\% \\
    CodeLlama-7b & 29.50\% & 32.14\% & 7.86\% & 30.87\% & 29.13\% & 0.37\% & 9.26\% & 25.11\% & 4.96\% \\
    DeepSeek-Coder-6b & 33.68\% & 36.02\% & 9.00\% & 21.36\% & 33.62\% & 0.43\% & 8.61\% & 24.96\% & 5.18\% \\
    \bottomrule
    \end{tabular}
    }
\end{table}

As presented in Table~\ref{tab:metrics_compirason}, all LLMs perform significantly worse on text-based metrics, with EM scores ranging from only 0.04\% to 0.55\%, a stark contrast to their performance on runtime-based metrics. 
In terms of BLEU and CodeBLEU scores, LLMs still underperform, with a BLEU score range from 5.56\% to 12.65\% and a CodeBLEU score range from 18.23\% to 25.11\%, remaining unexpectedly low compared to their performance on the other two tasks.
Despite the low EM scores, the generated code maintains a relatively low syntax error rate, ranging from 4.96\% to 15.47\%, similar to results observed on Defects4J. Interestingly, decoder-only models also perform well in syntax error rate, likely due to the Methods2Test dataset’s filtering process, which removes overly long inputs.

We attribute this lack of correlation to the diversity of the test cases. As mentioned in Section~\ref{sec:dataset}, half of the focal methods in the Methods2Test dataset include at least two related test cases, reflecting real-world development practices. For example, the focal method \texttt{`public Collection<Attribute> encode();'} in the Methods2Test dataset has eight associated test cases, each testing the method across different specific encoding types and verifying that the correct collection is returned. 
Although we remove duplicate test cases, the overall distribution remains unchanged. This may explain why LLMs perform poorly on text-based metrics yet relatively well on runtime-based metrics: the generated test cases may test the focal method from perspectives different from those in the ground truth.
These findings indicate that, given the diversity of test cases and the real-world development context, runtime-based metrics are more suitable than text-based metrics for evaluating the effectiveness of automated test generation approaches.

\summary{
LLMs may perform poorly on text-based metrics due to the diversity of test cases. Considering realistic development scenarios, it is recommended to use runtime-based metrics to evaluate the effectiveness of the automated test generation approaches.
}

\section{Guidelines}
\label{sec:implication}

We summarize some guidelines to suggest better practices for leveraging LLMs in unit testing.

\textbf{(1) Potential of LLM-based approaches.} 
As discussed in Section~\ref{sec:rq1}, our study shows that fine-tuning LLMs significantly outperforms state-of-the-art methods such as $ATLAS$ for assertion generation, \textsc{AthenaTest} for test generation, and \textsc{Ceprot} for test evolution across nearly all metrics. This improvement highlights the potential of LLMs for a wide range of unit testing tasks. Therefore, LLM-based approaches should be regarded as a primary strategy for achieving superior results.

\textbf{(2) Incorporating Additional Post-Processing.}
As detailed in Section~\ref{sec:rq1-test_generation}, LLMs 
fall short compared to A3Test, which, though also based on fine-tuning, incorporates a post-processing step to ensure naming consistency, parentheses completion, and test signature accuracy. This suggests that even a straightforward post-processing step can effectively enhance LLMs' performance. Thus, exploring post-processing techniques, especially those that integrate program analysis knowledge, may be essential to enhance LLM effectiveness further.

\textbf{(3) Importance of Input Length.}
In Section~\ref{sec:rq1-test-evolution}, we describe reallocating input space in the \textsc{Ceprot} approach by removing the edit sequence to include additional content—such as the original method, updated method, and original test. This adjustment led to a 4.7\% improvement in EM and a 24.1\% increase in CodeBLEU, demonstrating that extending input context length significantly enhances LLMs' performance in the future.

\textbf{(4) Selection of LLMs.}
As discussed in Section~\ref{sec:rq2}, we provide empirical recommendations for selecting LLMs in unit testing tasks. First, large-scale LLMs consistently outperform smaller ones, indicating that larger models should be prioritized when sufficient computing resources are available. The CodeLlama and DeepSeek-Coder series are especially worth considering for their strong performance. Second, encoder-decoder architectures perform better among models with fewer than one billion parameters. 
Thus, in resource-constrained settings, encoder-decoder models, particularly the CodeT5p series, maybe the optimal choice for unit testing tasks.

\textbf{(5) Potential Issue with Failing Tests and Build Errors.}
As illustrated in Fig.~\ref{fig:rq2_3_all_tasks}, a considerable proportion of generated test cases fail when executed against production code, regardless of model size.
Moreover, according to Table~\ref{tab:tg_comparisons_llm}, even the top-performing model, DeepSeek-Coder-6b, still suffers from a high failure and build error rate of 30.36\% on Defects4J.
These failures, often resulting from incorrect test prefixes or assertions, significantly hinder the effectiveness of automated test case generation and impair the precision of bug detection. 
Therefore, reducing failing and build error test rates is crucial for further advancements.

\textbf{(6) Runtime-based Metrics rather than Text-based Metrics.}
As discussed in Section~\ref{sec:metrics_comparison}, LLMs may perform poorly on text-based metrics due to the diversity of test cases, which significantly impacts test generation tasks. The other two tasks may also be affected. Therefore, it is recommended to use runtime-based metrics to evaluate approaches for unit testing tasks.

\section{Threats to Validity}
\label{sec:validity}

\textbf{Task Selection.}
Task selection is a critical factor affecting validity, as unit testing encompasses a wide domain with numerous tasks from diverse perspectives.
To mitigate this threat, we select three widely recognized unit testing tasks representing different granularities and scenarios.

\textbf{Data Leakage.} 
Data leakage poses another potential risk, particularly in the test generation task using Defects4J dataset as an evaluation benchmark. To address this issue, we evaluate the performance of LLMs on the internal dataset to validate the robustness of our conclusions.

\textbf{LLM Selection.} 
The final threat to validity is the choice of LLMs. 
A limited number or scale of models could impact the robustness of our conclusions. 
To mitigate this, we select 37 LLMs with diverse architectures and a wide range of sizes, enabling the largest study in the literature.

\sloppy
\section{Conclusion}
\label{sec:conclusion}

In this paper, we conduct a large-scale empirical study on fine-tuning LLMs for unit testing, involving 37 widely used LLMs and three unit testing tasks. 
Our results demonstrate that LLMs outperform state-of-the-art approaches across nearly all metrics, highlighting the substantial advancement of LLMs in unit testing tasks. 
Besides, we analyze the impact of various factors on the performance of LLMs, such as model architectures, and discuss some key topics, such as data leakage and bug detection.
Lastly, our findings provide various practical guidelines for future LLM-based unit testing research.
Overall, this work highlights the potential of LLMs to advance unit testing and provides valuable insights for researchers to design effective unit testing approaches in the future.

\section{Data Availability}
Our experimental materials are available at \url{https://github.com/iSEngLab/LLM4UT_Empirical}.

\section*{Acknowledgments}
The authors would like to thank the anonymous reviewers for their insightful comments.
This work is supported partially by the National Natural Science Foundation of China (61932012, 62372228), CCF-Huawei Populus Grove Fund (CCF-HuaweiSE202304, CCF-HuaweiSY202306) and Science, Technology and Innovation Commission of Shenzhen Municipality (CJGJZD20200617103001003).

\bibliographystyle{ACM-Reference-Format}
\bibliography{reference}

\end{document}